\documentclass[runningheads]{llncs}

{\bibliographystyle{splncs04}}

\usepackage[colorlinks=true,urlcolor=blue,citecolor=blue,linkcolor=blue]{hyperref}
\usepackage{multirow}    
\usepackage{pifont}      
\usepackage{color}       
\usepackage{xspace}      
\usepackage{flushend}     
\usepackage{graphicx}
\usepackage{caption}
\usepackage{subfigure}
\usepackage{url}
\usepackage{rotating}
\usepackage{wrapfig}
\usepackage[misc]{ifsym}

\begin{document}

\newcommand{\mypara}[1]{\vspace{2pt}\noindent\textbf{{#1. }}}
\newcommand{\name}{KASR\xspace}
\newcommand{\eat}[1]{}  

\newcommand{\authcomment}[3]{\textcolor{#3}{#1 says: #2}}
\newcommand{\zhi}[1]{\authcomment{zhi}{#1}{blue}}


\title{\name: A Reliable and Practical Approach to Attack Surface Reduction of \\ Commodity OS Kernels}

\titlerunning{A Reliable and Practical Approach to Kernel Attack Surface Reduction}

\author{Zhi Zhang\inst{1}\textsuperscript{,}\inst{2}\textsuperscript{(\Letter)} \and
Yueqiang Cheng\inst{3} \and
Surya Nepal\inst{1} \and
Dongxi Liu\inst{1} \and
Qingni Shen\inst{4} \and
Fethi Rabhi\inst{2} 
}
\authorrunning{Z. Zhang et al.}
%
\institute{Data61, CSIRO, Australia  
\email{\normalsize \{zhi.zhang,surya.nepal,dongxi.liu\}@data61.csiro.au} \and
University of New South Wales, Sydney, Australia \\ 
\email{\normalsize zhi.zhang@student.unsw.edu.au}, \email{\normalsize f.rabhi@unsw.edu.au} \and
Baidu XLab, Sunnyvale, California, United States \\
\email{\normalsize chengyueqiang@baidu.com} \and
Peking University, Beijing, China \\
\email{\normalsize qingnishen@ss.pku.edu.cn}}
%
%
%
%
%
%
%
%
%
%
%


\maketitle

\thispagestyle{empty}
\begin{abstract}
Commodity OS kernels have broad attack surfaces due to the large code base and the numerous features such as device drivers. For a real-world use case (e.g., an Apache Server), many kernel services are \emph{unused} and only a small amount of kernel code is \emph{used}. Within the \emph{used} code, a certain part is invoked only at runtime while the rest are executed at startup and/or shutdown phases in the kernel's lifetime run.

In this paper, we propose a \emph{reliable} and \emph{practical} system, named \name,
which \emph{transparently} reduces attack surfaces of commodity OS kernels at runtime without requiring their source code.
The \name system, residing in a trusted hypervisor, achieves the attack surface reduction through a two-step approach:
(1) reliably depriving \emph{unused} code of executable permissions,
and (2) transparently segmenting \emph{used} code and selectively activating them.
We implement a prototype of \name on Xen-$4.8.2$ hypervisor and evaluate its security effectiveness on Linux kernel-$4.4.0$-$87$-generic.  
Our evaluation shows that \name reduces the kernel attack surface by $64\%$ and trims off $40\%$ of CVE vulnerabilities.
Besides, \name successfully detects and blocks all $6$ real-world kernel rootkits. 
We measure its performance overhead with three benchmark tools (i.e., \texttt{SPECINT}, \texttt{httperf} and \texttt{bonnie++}). 
The experimental results indicate that \name imposes less than 1\% performance overhead (compared to an unmodified Xen hypervisor) on all the benchmarks.
\end{abstract}

\keywords{Kernel Attack Surface Reduction, Reliable and Practical System, Hardware-Assisted Virtualization}

%



\section{Introduction}\label{sec:introduction}
In order to satisfy various requirements from individuals to industries, commodity OS kernels have to support numerous features, including various file systems and numerous peripheral device drivers. These features inevitably result in a broad attack surface, and this attack surface becomes broader and broader with more services consolidated into the kernel every year.
As a consequence, the current kernel attack surface gives an adversary numerous chances to compromise the OS kernel and exploit the whole system.
Although we have moved into the virtualization and cloud era,
the security threats are not being addressed. Instead it becomes even worse with the introduction of additional software stacks, e.g., a hypervisor layer.
Recent years have witnessed many proposed approaches which realized the severity of this issue and made an effort to reduce the attack surface of the virtualized system. Specifically, schemes like NoHype~\cite{nohype-ccs}, XOAR~\cite{xoar-sosp} HyperLock~\cite{hyperlock-eurosys} and Min-V~\cite{min-v-eurosys} are able to significantly reduce the attack surface of the hypervisor. In addition, several other schemes have been proposed to reduce the huge kernel attack surface, which are summarized into the following three categories.

\eat{
Intuitively, for a stable \emph{Apache} server use case, many kernel functions are unneeded and the needed kernel functionality can be determined in advance. The unnecessary functions still reside in memory, contributing to a large portion of the kernel attack surface. If it can be minimized for the use case, the risk of kernel being compromised will be efficiently reduced. 
}


\eat{
Recent years have witnessed many profiling approaches. They have made an effort to minimize the kernel attack surface by locking the kernel services required by a specified workload while disabling the unnecessary. 
Specifically, Tartler~\cite{tartler2012automatic} and Kernel Tailoring~\cite{attacksurface-ndss} patch kernel configurations that is adapted to a particular workload and then recompile the kernel source code.  A common limitation is that recompling the source code is difficult to gain the distribution support. 
Instead of making any changes to the production systems, Face-Change~\cite{gu2014face} is a Virtual Machine Introspection (VMI)-based technique to profile the kernel for a target application. Based on the profiling results, it could provide a minimal kernel code base
for the application. However, it supports neither the Kernel level Address Space Layout Randomization (KASLR)~\cite{cook2013linux} nor multiple vCPU in a single Virtual Machine (VM). On top of that, it induces a worst-case overhead of $40$\%. Thus, these disadvantages impede its deployment in practice.
}




\textbf{Build from Scratch.} The first category attempts to build a micro-kernel with a minimal attack surface~\cite{accetta1986mach,herder2006construction,herder2006minix,sel4}, among which Sel4~\cite{sel4} is the first OS that achieves a high degree of assurance through formal verification.
Although such micro-kernel schemes retrofit security, they are incompatible with legacy applications.

\textbf{Re-Construction.} The second category makes changes to current monolithic kernel. Nooks~\cite{swift2002nooks}, and LXFI~\cite{LXFI2011software} isolate buggy device drivers to reduce the attack surface of the kernel. Considering that the reduced kernel is still large, Nested Kernel~\cite{nestedkernel-asplos} places a small isolated kernel inside the monolithic kernel, further reducing the attack surface. Besides, strict access-control policies~\cite{cook2013linux,selinux2001implementing} and system call restrictions~\cite{Seccomp} also contribute a lot. A common limitation of these approaches is that they all require modifications of the kernel source code, which is usually not applicable.

\textbf{Customization.} The last category manages to tailor existing kernels without modifications. 
Tartler~\cite{tartler2012automatic}, Kernel Tailoring~\cite{attacksurface-ndss} and Lock-in-Pop~\cite{li2017lock} require the Linux source code of either the kernel or core libraries (i.e., \emph{glibc}) to restrict user's access to the kernel. They lack the OS distribution support due to the requirement of source code re-compiling.
Ktrim~\cite{Ktrim} and KRAZOR~\cite{kurmus2014quantifiable} rely on specific kernel features (i.e., \emph{kprobes}) to binary-instrument kernel functions and remove unused ones.
Face-Change~\cite{gu2014face} is a hypervisor-based technique to tailor the kernel code. It supports neither the Kernel Address Space Layout Randomization (KASLR)~\cite{cook2013linux} nor multiple-vCPU for the target kernel. Besides, it induces a worst-case overhead of $40$\%, impeding its deployment in practice.

\textbf{Overview.} In this paper, we propose a \emph{reliable} and \emph{practical} virtualized system, named \name, which is able to \emph{transparently} reduce the attack surface of a commodity OS kernel at runtime.

%
\eat{
As discussed above, for a typical workload, 
only a subset of the kernel services are enough to support the workload. 
In contrast to the \emph{used} kernel code, many other kernel services are never executed. We call them \emph{unused} kernel code in this paper.
}

Consider a specified application workload (e.g., an Apache server), whose operations do not necessarily need all kernel services. Instead, only a subset of the services are invoked to support both the target Apache process and the kernel. For example, both of them always require code blocks related to memory management (e.g., \emph{kmalloc}, \emph{kfree}, \emph{get\_page}) and synchronization mechanisms (e.g., \emph{\_spin\_lock}).
Apart from that, certain \emph{used} kernel functions are only used during a specific period of kernel's lifetime and remain unused for the rest of the time. For instance,
the initialization (e.g., \emph{kernel\_init}) and power-off actions (e.g., \emph{kernel\_power\_off}) will only be taken when the kernel starts up and shuts down, respectively. 
In contrast to these \emph{used} kernel code, many other kernel services are never executed. We call them \emph{unused} kernel code in this paper. The unused kernel code resides in the main memory, contributing to a large portion of the kernel attack surface. For example, a typical kernel vulnerability, e.g., CVE-$2013$-$2094$, is exploited via a crafted system call \emph{perf\_event\_open} that is unused or never invoked in the Apache workload.

Motivated by the above observation, \name achieves the kernel attack surface reduction in two steps.
The first step is to reliably deprive unused code of executable permissions.
Commodity OS kernels are designed and implemented to support all kinds of use cases (e.g., the Apache server and Network File System service), and therefore there will be a large portion of kernel code (e.g., system call handlers) unused for a given use case. By doing so, this step could effectively reduce a large portion of the attack surface.
The second step transparently segments used code and selectively activates it according to the specific execution demands of the given use case.
This segmentation is inspired by the observation that certain kernel code blocks (e.g., \emph{kernel\_init}) only execute in a particular period, and never execute beyond that period.
As a result, \name dramatically reduces the attack surface of a running OS kernel. 

We implement a \name prototype on a private cloud platform, with Xen $4.8.2$ as the hypervisor and Ubuntu Server $16.04.3$ LTS as the commodity OS. The OS kernel is unmodified Linux version $4.4.0$-$87$-generic with KASLR~\cite{cook2013linux} enabled.
\name only adds about $1.2K$ SLoC to the hypervisor code base. 
We evaluate its security effectiveness under the given use cases (e.g., Linux, Apache, MySQL and PHP (LAMP)-based server).  
The experimental results indicate that \name reduces more than $64\%$ kernel attack surface at the granularity of code pages. 
Also, we trims off $40\%$ of Common Vulnerabilities and Exposures (CVEs), since the CVE reduction indicates the number of CVEs that \name could avoid. 
In addition, \name successfully detects and blocks all $6$ real-world kernel rootkits.
We also measure the performance overhead using several popular benchmark tools as given use cases, i.e., \texttt{SPECint}, \texttt{httperf} and \texttt{bonnie++}.
The overall performance overheads are $0.23\%$, $0.90\%$ and $0.49\%$ on average, respectively.

\mypara{Contributions}
In summary, we make the following key contributions:
\begin{itemize}
	\item Propose a novel two-step approach to reliably and practically reduce the kernel attack surface with being agnostic to the particular OS.
	\item Design and implement a practical \name system on a recent private cloud platform. \name transparently ``fingerprints'' used kernel code and enables them to execute according to their execution phases. 
	\item Evaluate the security effectiveness of the \name system by the reductions of kernel attack surface, CVE and the mitigation of real-world rootkits.
	\item Measure the performance overhead of the \name system using several popular benchmark tools. The low overhead makes \name reasonable for real-world deployment.
\end{itemize}

\mypara{Organization}
The rest of the paper is structured as follows.
In Section~\ref{sec:prob}, we briefly describe our system goals and a threat model.
In Section~\ref{sec:rationale}, we present the kernel attack surface, its measurement and the rationale of its reduction.
We introduce in detail the system architecture of \name in Section~\ref{sec:overview}.
Section~\ref{sec:impl} and Section~\ref{sec:eva} present the primary implementation of \name and its performance evaluation, respectively.
In Section~\ref{sec:dis} and Section~\ref{sec:related}, we discuss limitations of \name, 
and compare it with existing works, respectively.
At last, we conclude this paper in Section~\ref{sec:con}.

\section{Threat Model and Design Goals} \label{sec:prob}
Before we describe our design, we specify the threat model and the design goals. 

\subsection{Threat Model}
In this paper, we focus on reducing the attack surfaces of commodity OS kernels in a virtualized environment.
Currently, most personal computers, mobile phones and even embedded devices are armed with the virtualization techniques, such as Intel~\cite{intelvt}, AMD~\cite{amdvt} and ARM virtualization support~\cite{armvt}.
Thus, our system can work on such devices.

We assume a hypervisor or a Virtual Machine Monitor (VMM) working beneath the OS kernel.
The hypervisor is trusted and secure as the root of trust.
Although there are vulnerabilities for some existing hypervisors, we can leverage additional security services to enhance their integrity~\cite{hypersafe,guardian,hypersentry} and reduce their attack surfaces~\cite{nohype-ccs,xoar-sosp}. 
As our system relies on a training-based approach, we also assume the system is clean and trusted in the training stage, but it could be compromised at any time after that.

We consider threats coming from both remote adversaries and local adversaries.
A local adversary resides in user applications, such as browsers and email clients.
The kernel attack surface exposed to the local adversary includes system calls, exported virtual file system (e.g., Linux \emph{proc} file system) for user applications.
A remote adversary stays outside and communicates with the OS kernel via hardware interfaces, such as a NIC.
The kernel attack surface for the remote adversary usually refers to device drivers.


\subsection{Design Goals}\label{sec:goals}
Our goal is to design a reliable, transparent and efficient system to reduce the attack surfaces of commodity OS kernels.

\mypara{G1: Reliable}
The attack surface should be reliably and persistently reduced. 
Even if kernel rootkits can compromise the OS kernel, they cannot enlarge the reduced attack surface to facilitate subsequent attacks.


\mypara{G2: Transparent}
The system should transparently work for the commodity OS kernels. Particularly, it neither relies on the source code nor breaks the kernel code integrity through binary instrumentation.
Source code requirement is difficult to be adopt in practice. And breaking the code integrity raises compatibility issues against security mechanisms, such as Integrity Measurement Architecture.


\mypara{G3: Efficient}
The system should minimize the performance overhead, e.g., the overall performance overhead on average is less than $1\%$.

Among these goals, G1 is for security guarantee, while the other two goals (G2 and G3 ) are for making the system practical. 
Every existing approach has one or more weaknesses: they either are unreliable (e.g., Lock-in-Pop~\cite{li2017lock} as per G1), or depend on the source code (e.g., SeL4~\cite{sel4}),  or break the kernel code integrity (e.g., Ktrim~\cite{Ktrim}), or incur high performance overhead (e.g., Face-Change~\cite{gu2014face}).
Our \name system is able to achieve all the above goals at the same time.

\section{Design Rationale}\label{sec:rationale}
We first present how to measure the attack surface of a commodity OS kernel, and then illustrate how to reliably and practically reduce it.

\subsection{Attack Surface Measurement}\label{sec:atm}

To measure the kernel attack surface, we need a security metric that reflects the system security.
Generally, the attack surface of a kernel is measured by counting its source line of code (SLoC).
This metric is simple and widely used.
However, this metric takes into account all the source code of a kernel, regardless of whether it is effectively compiled into the kernel binary.
To provide a more accurate security measurement, Kurmus et al.~\cite{attacksurface-ndss}
propose a fine-grained generic metric, named GENSEC,
which only counts effective source code compiled into the kernel.
More precisely, in the GENSEC metric, the kernel attack surface is composed of the entire running kernel,
including all the Loadable Kernel Modules (LKMs).

However, the GENSEC metric only works with the kernel source code, rather than the kernel binary. 
Thus it is not suitable for a commodity OS with only a kernel binary that is made of a kernel image and numerous module binaries.
To fix this gap, we apply a new \name security metric.
Specifically, instead of counting source lines of code, the \name metric counts all executable instructions.

Similar to prior schemes that commonly use SLoC as the metric of the attack surface,
the \name metric uses the Number of Instructions (NoI).
It naturally works well with instruction sets where all the instructions have an equal length (e.g., ARM instructions).
However, with a variable-length instruction set (e.g., x86 instructions~\cite{intelvt}),
it is hard to count instructions accurately.
In order to address this issue on such platforms, we use the Number of Instruction Pages (NoIP).
NoIP is reasonable and accurate due to the following reasons.
First, it is consistent with the paging mechanism that is widely deployed by all commodity OS kernels.
Second, the kernel instructions are usually contiguous and organized in a page-aligned way.
Finally, it could smoothly address the issue introduced by variable-length instructions without introducing any explicit security and performance side-effects.
In this paper, the \name metric depends on NoIP to measure the kernel attack surface.

\subsection{Benefits of Hardware-assisted Virtualization}
In a hardware-assisted virtualization environment, there are two levels of page tables.
The first-level page table, i.e., Guest Page Table (GPT), is managed by the kernel in the guest space, and the other one, i.e., Extended Page Table (EPT), is managed by the hypervisor in the hypervisor space.
The hardware checks the access permissions at both levels for a memory access.
If the hypervisor removes the executable permission for a page $P_a$ in the EPT, then the page $P_a$ can never be executed, regardless of its access permissions in the GPT.
These mechanisms have been widely supported by hardware processors (e.g., Intel~\cite{intelvt}, AMD~\cite{amdvt}, and ARM~\cite{armvt}) and commodity OSes.

With the help of the EPT, we propose to reduce the attack surface by transparently removing the executable permissions of certain kernel code pages.
This approach achieves all system goals listed before.
First, it is reliable (achieving G1) since an adversary in the guest space does not have the capability of modifying the EPT configurations.
Second, the attack surface reduction is transparent (achieving G2), as the page-table based reduction is enforced in the hypervisor space, without requiring any modifications (e.g., instruction instrumentation) of the kernel binary.
Finally, it is efficient (achieving G3) as all instructions within pages that have executable permissions are able to execute at a native speed.

\section{\name Design} \label{sec:overview}

\begin{figure}
\centering
\includegraphics[height=4.5cm]{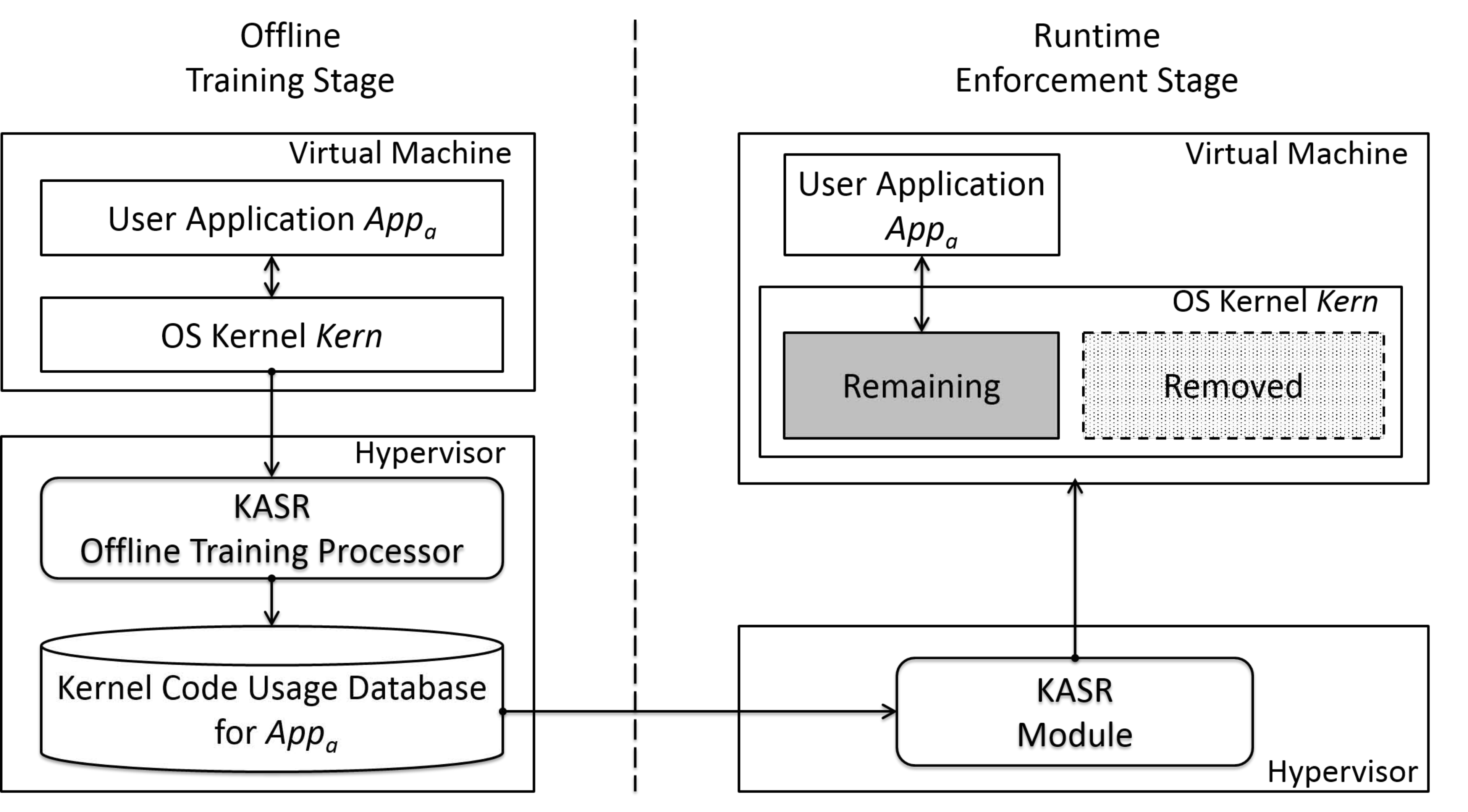} \\
\caption{\textbf{The architecture of the \name system.}} 
\label{fig:arch}
\end{figure}


We firstly elaborate the design of the \name system. As depicted in Figure~\ref{fig:arch}, the general working flow of \name proceeds in two stages: an offline training stage followed by a runtime enforcement stage. In the offline training stage, a trusted OS kernel \emph{Kern} is running beneath a use case (e.g., user application \emph{$App_a$}) within a virtual machine. The \name offline training processor residing in the hypervisor space, monitors the kernel's lifetime run, records its code usage and generates a corresponding database. 
The generated kernel code usage database is trusted, as the system in the offline training stage is clean.
Once the generated database becomes stable and ready to use, the offline training stage is done.

In the runtime enforcement stage, the \name module, running the same virtual machine, loads the generated database and reduces the attack surface of \emph{Kern}. 
The kernel attack surface is made up of the kernel code from the kernel image as well as loaded LKMs.
A large part of the kernel attack surface is reliably removed (the dotted square in Figure~\ref{fig:arch}). Still, the remaining part (the solid shaded-square in Figure~\ref{fig:arch}) is able to support the running of the use case $App_a$. 
The attack surface reduction is reliable, as the hypervisor can use the virtualization techniques to protect itself and the \name system, indicating that no code from the virtual machine can revert the enforcement.

\subsection{Offline Training Stage}
\begin{figure}
\centering
\includegraphics[height=4.0cm]{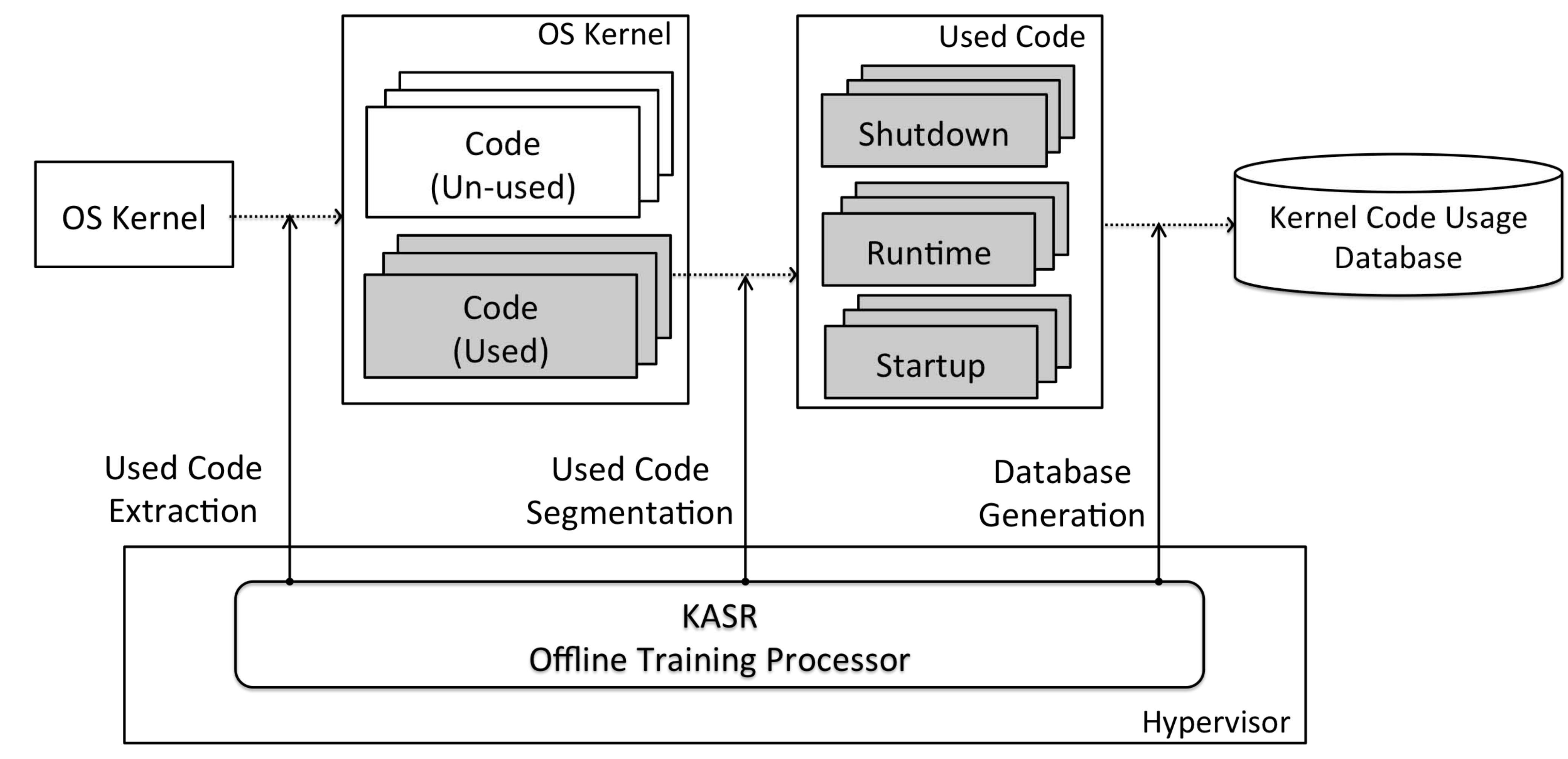} \\
\caption{\textbf{Offline Training Stage.} The \name offline training processor working in the hypervisor space, extracts used code from the OS kernel, segments used code into three phases (i.e., startup, runtime and shutdown) and generates the kernel code usage database.}
\label{fig:offlinearch}
\end{figure}

Commodity OSes are designed and implemented to support various use cases.
However, for a given use case (e.g., \emph{$App_a$}), only certain code pages within the kernel (e.g., \emph{Kern}) are used while other code pages are unused. 
Thus, the \name offline training processor can safely extract the used code pages from the whole kernel, the so-called used code extraction. 
On top of that, the used code pages can be segmented into three phases (e.g., startup, runtime and shutdown). 
The code segmentation technique is inspired by the observation that some used code pages are only used in a particular time period. 
For instance, the \emph{init} functions are only invoked when the kernel starts up and thus they should be in the \emph{startup} phase. 
However, for certain functions, e.g., \emph{kmalloc} and \emph{kfree}, they are used during the kernel's whole lifetime and owned by all three phases. 
The \name offline training processor uses the \emph{used code extraction} technique (Section~\ref{sec:uce}) to extract the used code pages, and leverages the \emph{used code segmentation} technique (Section~\ref{sec:ucs}) to segment used code into different phases.
All the recorded code usage information will be saved into the kernel code usage database, as shown in Figure~\ref{fig:offlinearch}.

The database will become stable quickly after the \name offline processor repeats the above steps several times. 
Actually, this observation has been successfully confirmed by some other research works~\cite{attacksurface-ndss,Ktrim}. 
For instance, for the use case of \emph{LAMP}, a typical \emph{httperf}~\cite{httperf} training of about ten minutes is sufficient to detect all required features, although the \emph{httperf} does not cover all possible paths.
This observation is reasonable due to the following two reasons.
First, people do not update the OS kernel frequently, and thus it will be stable within a relatively long period.
Second, although the user-level operations are complex and diverse, the invoked kernel services (e.g., system calls) are relatively stable, e.g., the kernel code that handles network packets and system files is constantly the same.

\subsubsection{Used Code Extraction}\label{sec:uce}
A key requirement of this technique is to collect \emph{all} used pages for a given workload.
It means that the collection should cover the whole lifetime of an OS kernel, from the very beginning of the startup phase to the last operation of the shutdown phase.
A straightforward solution is to use the trace service provided by the OS kernel.
For instance, the Linux kernel provides the \emph{ftrace} feature to trace the kernel-level function usage.
However, all existing integrated tracing schemes cannot cover the whole life cycle. 
For example, \emph{ftrace} always misses the code usage of the startup phase~\cite{attacksurface-ndss} before it is enabled.
Extending the trace feature requires modifying the kernel source code.
To avoid the modification and cover the whole life cycle of the OS kernel, we propose a hypervisor-based \name offline training processor. The offline training processor, working in the hypervisor space, starts to run before the kernel starts up and remains operational after the kernel shuts down. 

In the following, we will discuss how to trace and identify the used code pages in the kernel image and loaded LKMs.

\mypara{Kernel Image Tracing}
Before the kernel starts to run, the offline training processor removes the executable permissions of all code pages of the kernel image.
By doing so, every code execution within the kernel image will raise an exception, driving the control flow to the offline training processor.
In the hypervisor space, the offline training processor maintains the database recording the kernel code usage status. 
When getting an exception, the offline training processor updates the corresponding record, indicating that a kernel code page is used.
To avoid this kernel code page triggering any unnecessary exceptions later, the offline training processor sets it to executable.
As a result, only the newly executed kernel code pages raise exceptions and the kernel continues running, thus covering the lifetime used code pages of the kernel image.
Note that the offline training processor filters out the user-space code pages by checking where the exception occurs. (i.e., the value of Instruction Pointer (IP) register).

\mypara{Kernel Modules Tracing}
The above tracing mechanism works smoothly with the kernel image, but not with newly loaded LKMs.
All LKMs can be dynamically installed and uninstalled into/from memory at runtime, and the newly installed kernel modules may re-use the executable pages that have already been freed by other modules in order to load their code. Thus, their page contents have totally changed and they become new code pages that ought to be traced as well. 
If we follow the kernel tracing mechanism, such to-be-reused pages cannot be recorded into the database. Because these pages have been traced and the processor has set them to executable, they are unable to trigger any exceptions even when they are reused by other modules. 

To address this issue, we dictate that only the page currently causing the exception can gain the executable permission while other pages cannot.  
Specifically, when a page $P_a$ raises an exception, the offline training processor sets it to executable so that the kernel can proceed to next page $P_b$. Once $P_b$ raises the exception, it is set to executable while $P_a$ is set back to non-executable. Likewise, the offline training processor sets $P_b$ back to non-executable when another exception occurs. 
By doing so, pages like $P_a$ or $P_b$ can trigger new exceptions if they will be re-used by newly installed modules and thus all used code pages can be traced.
Obviously, this approach is also suitable for the kernel image tracing.  
 

\mypara{Page Identification}
The traced information is saved in the database, and the database reserves a unique identity for each code page.
It is relatively easy to identify all code pages of the kernel image when its address space layout is unique and constant every time the kernel starts up.
Thus, a Page Frame Number (PFN) could be used as the identification.
However, recent commodity OS kernels have already enabled the KASLR technology~\cite{cook2013linux} and thus the PFN of a code page is no longer constant.
Likewise, this issue also occurs with the kernel modules, whose pages are dynamically allocated at runtime, and each time the kernel may assign a different set of PFNs to the same kernel module.

A possible approach is to hash every page's content as its own identity.
It works for most of the code pages but will fail for the code pages which have instructions with dynamically determined opcodes, e.g., for the \emph{call} instruction, it needs an absolute address as its operand, and this address may be different each time, causing the failure of page identification.
Another alternative is to apply the fuzzy hash algorithm (e.g., ssdeep~\cite{kornblum2010fuzzy}) over a page and compute a similarity (expressed as a percentage) between two pages. e.g., if two pages have a similarity of over $60\%$, they are identical. However, such low similarity will introduce false positives, which can be exploited by attackers to prompt malicious pages for valid ones in the runtime enforcement stage. 

To address the issues, we propose a multi-hash-value approach.
In this offline training stage, we trace the kernel for multiple rounds (e.g., $10$ rounds) to collect all the used pages and dump the page content of each used page. Then we build a map of what bytes are constant and what bytes are dynamic in every used page. Each used page has multiple ranges and each range is made up of consecutive constant bytes. The ranges are separated by the dynamic bytes. 
Based on the map, we compute a hash value for every range. If and only if two pages have the same hash value for each range, they are identical. 
As a result, a page's identity is to hash everything within the page but the dynamic bytes. On top of that, we observe that the maximum byte-length of the consecutive dynamic bytes is $4$, making it hardly possible for attackers to replace the dynamic bytes with meaningful rogue ones. Relying on the approach, the risk of abusing the false positives is minimized.

\eat{
pick $N$ short code blocks out of a page, compute a hash value for each one of them and store all the values into an array as the page's identity. 
If $M$ out of $N$ (e.g., more than one half) hash values in two pages are matched, it indicates that they are the same page. 
False positives occur if the ratio of $M$ to $N$ is too low while the page identification may fail if the ratio is too high. 
Based on our experiments, we resolve these issues by choosing a proper ratio (i.e., $M/N$ = $80\%$). 
Note that the $N$ code blocks do not have to cover a whole page as long as they can identify a page.
Besides, if the $N$ code blocks' starting page offset is predefined, attackers may craft a malicious code page where the $N$ blocks remain intact and the rest blocks are attack payloads. In that case, the \name module will regard the page as a used kernel page within the code-usage database and further allow it to execute in the runtime enforcement stage (see Section~\ref{sec:runenforce}). In order to mitigate the attack, where to pick the $N$ code blocks within a page is a secret value, which users determine before the offline training stage starts. And then it is stored in the hypervisor space. 
}

\subsubsection{Used Code Segmentation}\label{sec:ucs}
This technique is used to segment the used code into several appropriate phases.  
By default, there are three phases: \emph{startup}, \emph{runtime}, and \emph{shutdown}, indicating which phases the used code have been executed in. 
When the kernel is executing within one particular phase out of the three, the offline training processor marks corresponding code pages with that phase. After the kernel finishes its execution, the offline training processor successfully marks all used code pages and saves their records into the database.
To be aware of the phase switch, the offline training processor captures the phase switch events.
For the switch between \emph{startup} and \emph{runtime}, we use the event when the first user application starts to run, 
while for the switch between \emph{runtime} and \emph{shutdown}, we choose the execution of the \emph{reboot} system call as the switch event. 


\subsection{Runtime Enforcement Stage}\label{sec:runenforce}

\begin{figure}
\centering
\includegraphics[height=4.0cm]{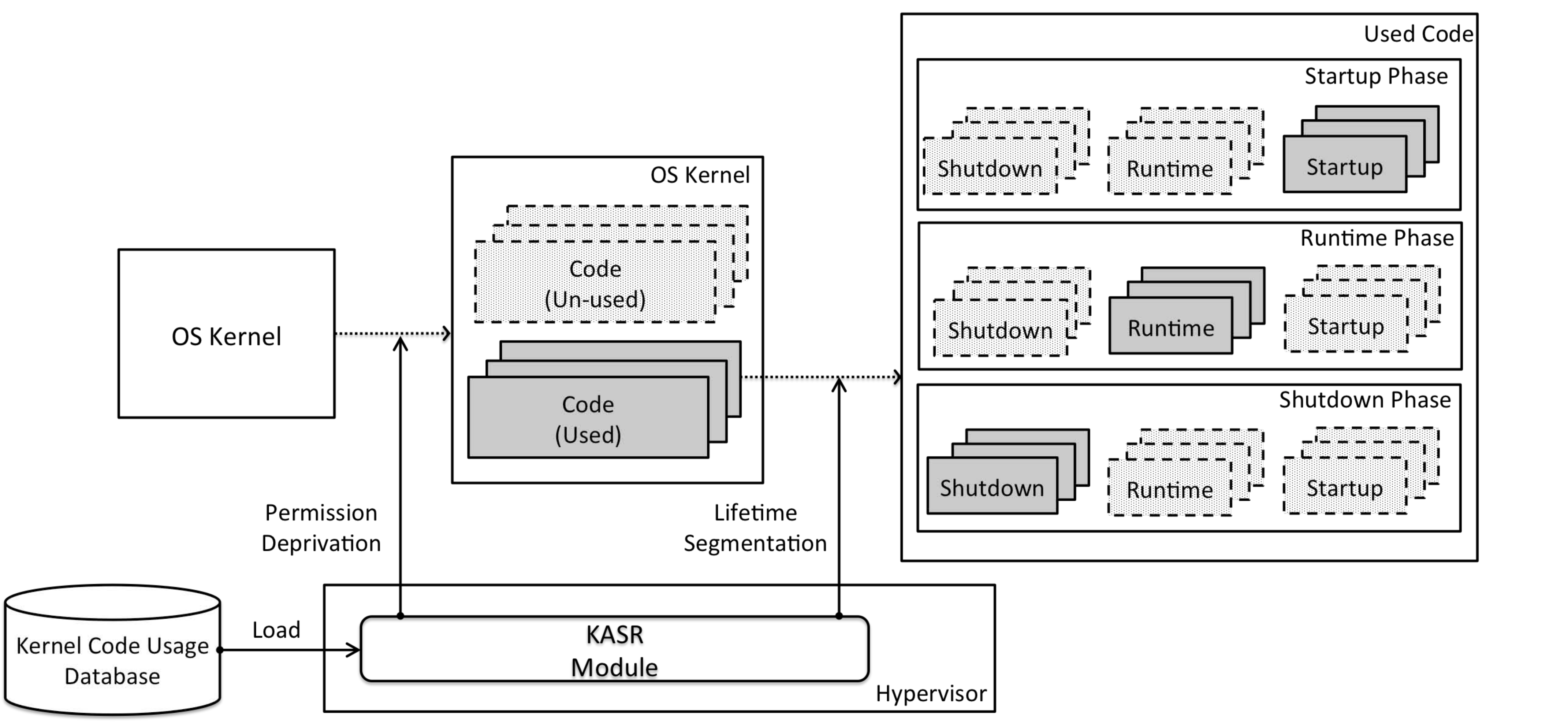} \\
\caption{\textbf{Runtime Enforcement Stage.} The \name module residing in the hypervisor space reduces OS kernel attack surface in two consecutive steps. The first step (i.e., permission deprivation) reliably deprives unused code of executable permission, and the second step (i.e., lifetime segmentation) selectively activates corresponding used code according to their phases.}
\label{fig:runtime-arch}
\end{figure}

When the offline training stage is done and a stable database has been generated (see details in Section~\ref{sec:dbbuild}), \name is ready for runtime enforcement.
As shown in Figure~\ref{fig:runtime-arch}, the \name module loads the generated database for a specific workload, and reduces the kernel attack surface in two steps:
\begin{enumerate}
\item \emph{Permission Deprivation.} It keeps the executable permissions of all used code pages (the solid shaded square in Figure~\ref{fig:runtime-arch}), and reliably removes the executable permissions of all unused code pages (the dotted square in Figure~\ref{fig:runtime-arch})
\item \emph{Lifetime Segmentation.} It aims to further reduce the kernel attack surface upon the permission deprivation. As shown in Figure~\ref{fig:runtime-arch}, it transparently allows the used kernel code pages of a particular phase to execute while setting the remaining pages to non-executable.
\end{enumerate}
All instructions within the executable pages can execute at a native speed, without any interventions from the \name module.
When the execution enters the next phase, the \name module needs to revoke the executable permissions from the pages of the current phase, and set executable permissions to the pages of the next phase.
To reduce the switch cost, the \name module performs two optimizations.
First, if a page is executable within the successive phase, the \name module skips its permission-revocation and keeps it executable.
Second, the \name module updates the page permissions in batch, rather than updating them individually. 

\eat{
\subsection{\name on System Call Optimization} 
As system calls are the major runtime interface that an adversary may utilize to launch attacks against the OS kernel,
it is meaningful for the \name system to have a particular optimization on system calls. 
The optimization aims to completely remove all unused system call handlers and selectively enable the used handlers according to the phase switch above.
However, the previous two techniques are page-based and cannot be directly applied to the system call Optimization, because the unused and used system call handlers may coexist on the same code pages.

Fortunately, all system call entries are located in a system call table, which is easy to manipulate. 
The \name system extracts all entries of the system calls from this table, removes unused ones and segments used ones into different phases.
Specifically, in the offline training stage, the offline training processor records the entries of all pages.
If one entry occurs, the offline training processor marks the corresponding system call as used.
Finally, the offline training processor produces the system call usage.
In the runtime enforcement stage, the \name module relies on the usage information to generate fake system call tables - retaining all used system call entries and removing the others.
When the phase switches (e.g., from the init phase to the runtime phase), the \name module switches the corresponding system call table accordingly.
This optimization is efficient (i.e., does not introduce extra performance overhead to the \name system) and effective (i.e., achieves a significant reduction of the system-call interface). 
}

\section{\name Database} \label{sec:impl}
This section presents the implementation details of the \name database, including database data-structure, database operations.

\subsection{Data Structure}

Basically, the database consists of two single-linked lists, which are used to manage the pages of kernel image and loaded modules, respectively. 
Both lists have their own list lock to support concurrent updates.
Every node of each list representing a page is composed of a node lock, a page ID, a status flag and a node pointer pointing to its next node.
The node lock is required to avoid race conditions and thus other nodes can be processed in parallel.

\mypara{Page ID}
The page ID is used to identify a page especially during the database updates. 
As kernel-level randomization is enabled within the kernel, we use the multi-hash-value approach for the identification.
Specifically, we trace the kernel for $10$ rounds to make sure that all the used pages are collected. Pages in different rounds are considered to be identical (i.e., a same page) if they satisfy two properties: (1) more than $3366$ out of $4096$ bytes (i.e., over $82\%$) are constant and the same among these pages; (2) the maximum byte-length of the consecutive different bytes (i.e., dynamic bytes) among these pages is no greater than $4$. And then we perform a per-byte comparison of the identical pages so as to build a map of what bytes are constant and what bytes are dynamic with the pages. By doing so, each used page has multiple ranges of consecutive constant bytes and dynamic bytes are between these ranges. As a result, all the constant bytes of every range are hashed as a value and all the hash values make up the page ID. 


\mypara{Status Flag}
The status flag indicates the phase status (i.e., \emph{startup}, \emph{runtime} and \emph{shutdown}) of a used page.
The flag is initialized as startup when the kernel boots up.
Once the kernel switches from the startup phase to the runtime phase, or from the runtime phase to the shutdown phase, appropriate exceptions are triggered so that the offline training processor can update the flag accordingly.
In our implementation, all code pages of the guest OS are deprived of executable permissions. Once the OS starts to boot, it will raise numerous EPT exceptions. In the hypervisor space, there is a handler (i.e., \emph{ept\_handle\_violation}) responding to the exception, and thus the offline training processor can mark the beginning of the runtime phase by intercepting the first execution of the user-space code as well as its end by intercepting the execution of the \emph{reboot} system call. 


\eat{
\mypara{Page ID}
In our implementation, the generic OS kernel does not support kernel-level randomization by default, and thus it is relatively simple to identify a code page of the kernel image.
Specifically, the base-address range of the kernel image remains the same every time the kernel starts up, from $0xc1000000$ to $0xc192e000$.
The relationship between linear addresses and physical addresses are constant one-to-one mappings.
Thus, .

For the kernel modules, every time they are installed, their base addresses are different, making a relative offset of instructions such as \emph{call} and \emph{jump} variable. As a result, these varied opcodes make the page content different each time even for the same page.
To choose proper page IDs for such code pages, we use the multi-hash-value based page ID. 
For the page ID, there are multiple hash values and each value corresponds to a specific range of consecutive stable bytes.

\mypara{Status Flag}
The status flag uses three bits to reserve three values ($1$, $2$, $4$), corresponding to the \emph{startup}, \emph{runtime} and \emph{shutdown} phases.

}
%

\subsection{Database Operations} \label{sec:dbbuild}
The database operations are mainly composed of three parts, i.e., populating, saving and loading.

\mypara{Populate Database}
To populate the database, the \name offline training processor must trace all the used pages and thus dictates that only the page raising the exception would become executable while others are non-executable.  
However, we find that this will halt the kernel.
The reason is that the x86 instructions have variable lengths and an instruction may cross a page boundary, which means that the first part of the instruction is at the end of a page, while the rest is in the beginning of the next page.
Under such situations, the instruction-fetch will result in infinite loops (i.e., trap-and-resume loops).

To address this issue, we relax the dictation and implement a queue of $2$ pages that own executable permissions. When the queue is full of two pages that have caused the first two exceptions (i.e., the first two used pages), it will then be updated by \emph{First-in, First-out}, i.e., the newest used page will be pushed in while the oldest used page will be popped out.
Besides solving the cross-page-boundary problem, we also accelerate the tracing performance.
Besides, we can capture all loaded modules, as all of them have no less than $2$ code pages.

To the end, it is not enough to obtain all the used pages by running the offline training stage just once.
Thus, it is necessary to repeat this stage for multiple rounds until the database size becomes stable.
In our experiments, $10$ rounds are enough to get a stable database (see Section~\ref{sec:eva}).

\mypara{Save and Load Database} \label{sec:dbrun}
The database is generated in the hypervisor space, and stored in the hard disk for reuse.
\eat{
There are two possible ways to save and load the database.
The straightforward approach is to add a disk and file system driver in the hypervisor to allow \name to directly write/read the database, which certainly adds hundreds of extra SLoC into the hypervisor space and might cause potential vulnerabilities.
To minimize the size of the hypervisor, the way 
, which eliminates the unnecessary security risks and is more flexible. 
We re-use the existing kernel drivers and export interfaces in the privileged domain (e.g., domain 0 of Xen). 
}
Specifically, we have developed a tiny tool in the privileged domain to explicitly save the database into the domain's disk after the offline training stage, and load the existing database into the hypervisor space during the runtime enforcement stage.

\section{Evaluation} \label{sec:eva} 
We have implemented a \name prototype on our private cloud platform, 
which has a Dell Precision T$5500$ PC with eight CPU cores (i.e., Intel Core Xeon-E$5620$) running at $2.40$\emph{GHz}. 
Besides, Intel VT-x feature is enabled and supports the page size of $4$\emph{KB}. 
Xen version $4.8.2$ is the hypervisor while Hardware-assisted Virtual Machine (HVM) is the Ubuntu Server $16.04.3$ LTS, which has a KASLR-enabled Linux kernel of version $4.4.0$-$87$-generic with four virtual CPU cores and $4$\emph{GB} physical memory. 
\name only adds around $1.2$\emph{K} SLoC in Xen.

In the rest of this section, we measure the reduction rates of the kernel attack surface.
On top of that, we characterize the reduced kernel attack surface in the metrics of Common Vulnerabilities and Exposures (CVEs).
The use cases we choose are \texttt{SPECint}, \texttt{httperf}, \texttt{bonnie++}, \texttt{LAMP} (i.e., Linux, Apache, MySQL and PHP) and \texttt{NFS} (i.e., Network File System). 
Furthermore, we test and analyze its effectiveness in defending against $6$ real-world kernel rootkits. 
Also, we measure the performance overhead introduced by \name through the selected use cases above. 
The experimental results demonstrate that 
we can effectively reduce kernel attack surface by $64\%$, CVEs by $40\%$, 
safeguard the kernel against $6$ popular kernel rootkits and impose negligible (less than $1\%$) performance overhead on all use cases.

\subsection{Kernel Attack Surface Reduction} \label{ssec:kasrd}
In the runtime enforcement stage, we measure the kernel attack surface reduction through three representative benchmark tools, namely, \texttt{SPECint}, \texttt{httperf} and \texttt{bonnie++} and two real-world use cases (i.e., \texttt{LAMP} and \texttt{NFS}).

\texttt{SPECint}~\cite{specint} is an industry standard benchmark intended for measuring the performance of the CPU and memory. In our experiment, the tool has $12$ sub-benchmarks in total and they are all invoked with a specified configuration file (i.e., \emph{linux64-ia32-gcc43+.cfg}).

On top of that, we measure the network I/O of HVM using \texttt{httperf}~\cite{httperf}. HVM runs an Apache Web server and Dom$0$ tests its I/O performance at a rate of starting from $5$ to $60$ requests per second ($100$ connections in total).

Also, we test the disk I/O of the guest by running \texttt{bonnie++}~\cite{bonnie} with its default parameters. For instance, \texttt{bonnie++} by default creates a file in a specified directory, size of which is twice the size of memory.

Besides, we run the \texttt{LAMP}-based web server inside the HVM. Firstly, we use the standard benchmark \emph{ApacheBench} to continuously access a static PHP-based website for five minutes. And then a Web server scanner \emph{Nikto}~\cite{Nikto} starts to run so as to test the Web server for insecure files and outdated server software and also perform generic and server type specific checks. This is followed by launching \emph{Skipfish}~\cite{Skipfish}, an active web application security reconnaissance tool. It operates in an extensive brute-force mode to carry out comprehensive security checks. Running these tools in the LAMP server aims to cover as many kernel code paths as possible.

Lastly, the other comprehensive application is \texttt{NFS}. HVM is configured to export a shared directory via \texttt{NFS}. In order to stress the \texttt{NFS} service, we also use \texttt{bonnie++} to issue read and write-access to the directory.

\begin{table}
\footnotesize
\centering
\caption{In every case, the kernel code pages are significantly tailored after each step. Generally, \name can reduce the kernel attack surface by $54$\% after the permission deprivation, and $64$\% after the lifetime segmentation. (Orig.Kern = Original Kernel, Aft.Per.Dep. = After Permission Deprivation, Aft.Lif.Seg. = After Lifetime Segmentation)}
\begin{tabular}{lccccccccc}
\hline
\multirow{2}{*}{\textbf{Cases}}  & \textbf{Orig.Kern} & \multicolumn{4}{c}{\textbf{Aft.Per.Dep.}} & \multicolumn{4}{c}{\textbf{Aft.Lif.Seg.}} \\
 & Page(\#) & Page(\#) & & Reduction(\%) & & & Page(\#) & & Reduction(\%) \\ \hline
\multirow{1}{*}{\textbf{\texttt{SPECint}}} & $2227$ & $1034$ & & $54$\% & & & $808$ & & $64$\% \\ \hline 
\multirow{1}{*}{\textbf{\texttt{httperf}}} & $2236$ & $1026$ & & $54$\% & & & $763$ & & $66$\% \\ \hline 
\multirow{1}{*}{\textbf{\texttt{bonnie++}}} & $2235$ & $1034$ & & $54$\% & & & $761$ & & $66$\% \\ \hline 
\multirow{1}{*}{\textbf{\texttt{LAMP}}} & $2238$ & $1043$ & & $53$\% & & & $817$ &  & $63$\% \\ \hline 
\multirow{1}{*}{\textbf{\texttt{NFS}}} & $2395$ & $1096$ & & $54$\% & & & $939$ &  & $61$\% \\ \hline 
\end{tabular}
\label{tab:cases}
\end{table}

All results are displayed in Table~\ref{tab:cases}. Note that the average results for \texttt{SPECint} are computed based on $12$ sub-benchmark tools.
We determine two interesting properties of the kernel attack surface from this table.
First, the attack surface reduction after each step is quite significant and stable for different use cases.
Generally, the attack surface is reduced by roughly $54$\% and $64$\% after the permission deprivation and lifetime segmentation, respectively, indicating that less than half of the kernel code is enough to serve all provided use cases.
Second, complicated applications (i.e., \texttt{LAMP} and \texttt{NFS}) occupy more kernel code pages than the benchmarks, indicating that they have invoked more kernel functions.

\eat{
\begin{wrapfigure}{r}{6.5cm}
\includegraphics[width=6.5cm, height=4.0cm]{image/cve-syscall-reduction.png} \\
\caption{\textbf{CVE and Syscall Reduction.} In all use cases, \name reduces the average numbers of CVEs and Syscalls by respective $40$\% and $66$\%.}
\label{fig:cve-syscall-reduction}
\end{wrapfigure}
}

\subsubsection{CVE Reduction}
Although some kernel functions (e.g., architecture-specific code) contain past CVE vulnerabilities, they are never loaded into memory during the kernel's lifetime run and do not contribute to the attack surface. 
As a result, we only consider the CVE-vulnerable functions that are loaded into the kernel memory.
We investigate CVE bugs of recent two years that provide a link to the GIT repository commit and identify $14$ CVEs that exist in the kernel memory of all five use cases. 

We observe that \name has removed $40$\% of CVEs in the memory. To be specific, some CVE-vulnerable kernel functions within the unused kernel code pages are deprived of executable permissions after the permission deprivation. For example, the \emph{ecryptfs\_privileged\_open} function in CVE-2016-1583 before Linux kernel-$4.6.3$ is unused, thus being eliminated. After the lifetime segmentation, some other vulnerable functions are also removed (e.g., \emph{icmp6\_send} in CVE-$2016$-$9919$). 

\eat{
\subsubsection{System Call Reduction}
\name has successfully trimmed more than half (i.e., $66\%$ on average) of the system calls (abbreviated to syscalls in Linux) for all use cases. \name uses the virtual machine introspection techniques to locate the syscall table. The original number of syscalls located in the syscall table is $377$ in total. After \name is enabled, an average of $249$ syscalls is removed at runtime. 

Specifically, the permission deprivation eliminates all unused syscalls. It removes non-standard Linux extension syscalls (e.g., \emph{process\_vm\_readv}), obsolete syscalls (e.g., \emph{olduname}), and rarely used syscalls (e.g., \emph{vm86}).

After the lifetime segmentation, syscalls used in both startup and shutdown phases are removed. One representative syscall in the startup phase is \emph{chroot}, which sets the root directory at the beginning of the system bootup. While in the shutdown phase, one familiar syscall is \emph{reboot} which restarts or powers off the system. 
}
\eat{
\subsubsection{Device Driver Reduction} 
As device drivers are more buggy than the core kernel~\cite{swift2002nooks}, we also perform measurements for them alone in every use case. 
Device drivers~\cite{kadav2012understanding} are categorized into four classes, (i.e., \emph{Char}, \emph{Block}, \emph{Net} and \emph{Others}). 
The class \emph{Others} refers to particular drivers such as virtual device drivers of Xen within the HVM.
All drivers are located in the directory of /lib/modules/$4.4.0$-$87$-generic and they are allowed to be loaded into memory in an on-demand way. 
\name dictates that only drivers that have been traced are permitted to execute while the other drivers are strictly prohibited.

In Table~\ref{tab:driv_dis}, the device driver modules (\#) and their related CVE number (\#) of every sub-class are listed in two groups. Take the class \emph{Char} as an example, the original group has $1643$ driver modules which contain $13$ CVEs. And the \name group has only $1$ CVE included by $31$ driver modules.
The original group indicates that all the drivers on disk can be loaded into memory, while the \name group allows only a certain number of drivers to have access to memory.
As shown in Table~\ref{tab:driv_dis}, 
$54$ loaded drivers in total in the \name group, in every use case, only account for $1$\% of that (i.e., $3710$) in the original group, indicating that $99$\% drivers are unnecessary and never invoked. Correspondingly, $91$\% (i.e., $20$ out of $22$) CVEs included by the drivers cannot be triggered. 

\begin{table*}
\footnotesize
\centering
\caption{In all cases, Original means all the driver modules on disk can be loaded into memory while \name only allows $54$ out of $3710$ driver modules to be loaded and reduces the number of CVE from $22$ to $2$. The reduction rates of driver modules and CVEs are $99\%$ and $91\%$, respectively.}

\begin{tabular}{lcccccc}
\hline
\multirow{2}{*}{\textbf{Drivers \quad}} & 
\multicolumn{2}{c}{\textbf{\quad Original \quad}} & \multicolumn{2}{c}{\textbf{\quad \name \quad}} & \multicolumn{2}{c}{\multirow{2}{*}{\textbf{\quad Reduction \quad}}} \\ 
 & Mod(\#) \qquad & CVE(\#) \qquad & Mod(\#) \qquad & CVE(\#) \qquad &  & \\ \hline 
\multirow{1}{*}{\textbf{\texttt{Char}}} & $1643$ & $13$ & $31$ & $1$ &\quad $98\%$ \qquad & $92\%$ \\ \hline
\multirow{1}{*}{\textbf{\texttt{Block}}} & $328$ & $2$ & $12$ & $1$ &\quad $96\%$ \qquad & $50\%$ \\ \hline
\multirow{1}{*}{\textbf{\texttt{Net}}} & $517$ & $6$ & $9$ & $0$ &\quad $98\%$ \qquad & $100\%$ \\ \hline 
\multirow{1}{*}{\textbf{\texttt{Others}}} & $1222$ & $1$ & $2$ & $0$ &\quad $99\%$ \qquad & $100\%$ \\ \hline 
\multirow{1}{*}{\textbf{\texttt{Total}}} & $3710$ & $22$ & $54$ & $2$ &\quad $99\%$ \qquad & $91\%$ \\ \hline
\end{tabular}
\label{tab:driv_dis}
\end{table*}
}

\subsection{Rootkit Prevention} 
%

Even though the kernel attack surface is largely reduced by \name, still there may exist vulnerabilities in the kernel, which could be exploited by rootkits. 
We demonstrate the effectiveness of \name in defending against real-world kernel rootkits.
Specifically, we have selected $6$ popular real-world kernel rootkits coming from a previous work~\cite{NICKLE} and the Internet. These rootkits work on typical Linux kernel versions ranging from $3.x$ to $4.x$, representing the state-of-the-art kernel rootkit techniques. 
All these rootkits launch attacks by inserting a loadable module and they can be divided into three steps:
\begin{enumerate}
\item inject malicious code into kernel allocated memory; 
\item hook the code on target kernel functions (e.g., original syscalls); 
\item transfer kernel execution flow to the code. 
\end{enumerate}
\name is able to prevent the third step from being executed. Specifically, rootkits could succeed at Step-1 and Step-2, since they can utilize exposed vulnerabilities to modify critical kernel data structures, inject their code and perform target-function hooking so as to redirect the execution flow. However, they cannot execute the code in Step-3, because \name decides whether a kernel page has an executable permission. Recall that \name reliably dictates that unused kernel code (i.e., no record in the database) has no right to execute in the kernel space, including the run-time code injected by rootkits. Therefore, when the injected code starts to run in Step-3, EPT violations definitely will occur and then be caught by \name. 
The experimental results from Table~\ref{tab:rootkit} clearly show that \name has effectively defended against all $6$ rootkits. As a result, \name is able to defend against the kernel rootkits to a great extent.

\eat{
In fact, \name is able to prevent unauthorized code from running and all the above rootkits have to inject malicious (unauthorized) code to launch attacks.
Note that hardware-assisted VMBR (abbreviated for Virtual-Machine Based Rootkits) such as Blue Pill~\cite{rutkowska2006introducing} and Vitriol~\cite{zovi2006hardware} are no exceptions, because such rootkits still require unauthorized loadable modules to load and execute. As a result, \name is able to defend against all (known and unknown) kernel rootkits that rely on code-injection attacks.
}


\begin{table}
\footnotesize
\centering
\caption{\name successfully defended against all $6$ kernel rootkits. (LKM = Loadable Kernel Module)}
\begin{tabular}{cccc}
\hline
{\textbf{OS Kernel \quad}} & {\textbf{Rootkit}} & {\textbf{Attack Vector}} & {\textbf{Attack Failed?}}\\ \hline
\multirow{6}{*}{Linux $3.x$-$4.x$ \quad} & adore-ng & LKM & $\surd$ \\
             & xingyiquan & LKM & $\surd$ \\ 
			 & rkduck & LKM & $\surd$ \\ 
			 & Diamorphine & LKM & $\surd$ \\ 
			 & suterusu & LKM & $\surd$ \\
			 & nurupo & LKM & $\surd$ \\ 
\hline
\end{tabular}
\label{tab:rootkit}
\end{table}

\subsection{Performance Evaluation}\label{sec:pe}
In this section, we evaluate the performance impacts of \name on CPU computation, network I/O and disk I/O using the same settings as we measure the kernel attack surface reduction.
Benchmark tools are conducted with two groups, i.e., one is called Original (HVM with an unmodified Xen), the other is \name.

Specifically, \texttt{SPECint} has 12 sub-programs and the CPU overhead caused by \name within each sub-program is quite small and stable. In particular, the maximum performance overhead is $1.47$\% while the average performance overhead is $0.23$\% for the overall system.

\texttt{Httperf} tests the Apache Web server inside the HVM using different request rates. Compared to the Original, the network I/O overhead introduced by \name ranges from $0.00$\% to $1.94$\% and the average is only $0.90$\%.

The disk I/O results are generated by \texttt{bonnie++} based on two test settings, i.e., \emph{sequential input} and \emph{sequential output}. For each setting, the \emph{read}, \emph{write} and \emph{rewrite} operations are performed and their results indicate that \name only incurs a loss of $0.49$\% on average.

\eat{
\begin{table}
\vspace{-0.8cm} 
\setlength{\abovecaptionskip}{0.1cm}
\footnotesize
\centering
\caption{CPU computation Benchmark Results from \texttt{SPECint}. The average performance overhead is $0.23$\%.}
\begin{tabular}{llll}
\hline
{\textbf{Programs \quad} \quad} & {\textbf{Original ($s$)} \quad} & {\textbf{\name ($s$)} \quad} & {\textbf{Overhead} \quad} \\  \hline
{perlbench} & {$459$} & {$459$} & {$0.00$\%}\\
{bzip2} & {$826$} & {$824$} & {$-0.20$\%} \\
{gcc} & {$387$} & {$390$} & {$0.68$\%}\\
{mcf} & {$312$}  & {$313$} & {$0.46$\%} \\
{gobmk} & {$601$} & {$600$} & {$-0.22$\%} \\
{hmmer} & {$702$} & {$704$} & {$0.23$\%} \\
{sjeng} & {$690$} & {$690$} & {$0.00$\%} \\
{libquantum} & {$939$} & {$949$} & {$1.03$\%} \\
{h264ref} & {$1016$} & {$1016$} & {$0.00$\%} \\
{omnetpp} & {$324$} & {$323$} & {$-0.31$\%} \\
{astar} & {$551$} & {$559$} & {$1.47$\%} \\
{xalancbmk} & {$297$} & {$296$} & {$-0.41$\%} \\ \hline
{Average} &  &  & {$0.23$\%} \\ \hline
\end{tabular}
\label{tab:spec}
\end{table}

\begin{table}
\footnotesize
\centering
\caption{Network I/O Benchmark Results from \texttt{httperf}. The average performance overhead is $0.90$\%.}
\begin{tabular}{cccc}
\hline
\textbf{Request} \quad & {\textbf{Original} \quad} & {\quad \textbf{\name}} & \multirow{2}{*}{\textbf{Overhead}} \\
\textbf{Rate} \quad & \textbf{Net I/O} ($KB/s$)  \quad & \quad \textbf{Net I/O} ($KB/s$) \quad & \\ 
\hline
{$25$} & {$12.9$} & {$12.7$} & {$1.55$\%} \\
{$30$} & {$15.5$} & {$15.3$} & {$1.29$\%} \\
{$35$} & {$18.1$} & {$18.0$} & {$0.55$\%} \\
{$40$} & {$20.7$} & {$20.6$} & {$0.48$\%} \\
{$45$} & {$23.3$} & {$23.2$} & {$0.43$\%} \\
{$50$} & {$25.8$} & {$25.7$} & {$0.39$\%} \\
{$55$} & {$28.4$} & {$28.4$} & {$0.00$\%} \\
{$60$} & {$31.0$} & {$30.4$} & {$1.94$\%} \\
{$65$} & {$33.6$} & {$33.1$} & {$1.49$\%} \\
\hline
{Average} &  &  & $0.90$\% \\
\hline
\end{tabular}
\label{tab:httperf}
\end{table}

\begin{table}
\footnotesize
\centering
\caption{Disk I/O Benchmark Results from \texttt{bonnie++}. The average performance overhead is $0.49$\%.}
\begin{tabular}{ccccc}
\hline
\multicolumn{2}{c}{\multirow{1}{*}{\textbf{Operations} \quad}} & {\quad \textbf{Original} ($K/s$)} & {\quad \textbf{\name} ($K/s$)} & \multirow{1}{*}{\quad \textbf{Overhead}} \\
\hline
\multicolumn{2}{l}{Sequential Output} &\quad \multirow{2}{*}{$38187$} &\quad \multirow{2}{*}{$37896$} &\quad \multirow{2}{*}{$0.76$\%} \\
\multicolumn{2}{l}{per Character} & & & \\
\hline
\multicolumn{2}{l}{Sequential Output} &\quad \multirow{2}{*}{$36554$} &\quad \multirow{2}{*}{$35760$} &\quad \multirow{2}{*}{$2.17$\%} \\
\multicolumn{2}{l}{per Block} & & & \\
\hline
\multicolumn{2}{l}{Sequential Output} &\quad \multirow{2}{*}{$39686$} &\quad \multirow{2}{*}{$40242$} &\quad \multirow{2}{*}{$-1.4$\%} \\
\multicolumn{2}{l}{Rewrite} & & & \\
\hline
\multicolumn{2}{l}{Sequential Input} &\quad \multirow{2}{*}{$68251$} &\quad \multirow{2}{*}{$68146$} &\quad \multirow{2}{*}{$0.15$\%} \\
\multicolumn{2}{l}{per Character} & & & \\
\hline
\multicolumn{2}{l}{Sequential Input} &\quad \multirow{2}{*}{$1309993$} &\quad \multirow{2}{*}{$1299930$} &\quad \multirow{2}{*}{$0.77$\%} \\
\multicolumn{2}{l}{per Block} & & & \\
\hline
\multicolumn{2}{l}{Average} &   &   &\quad  $0.49$\% \\
\hline
\end{tabular}
\label{tab:bonnie}
\end{table}
}

\subsection{Offline Training Efficiency}
\begin{figure}[t]
\begin{minipage}[t]{0.49\linewidth}
\centering
\includegraphics[width=0.99\linewidth, height=0.3\textheight]{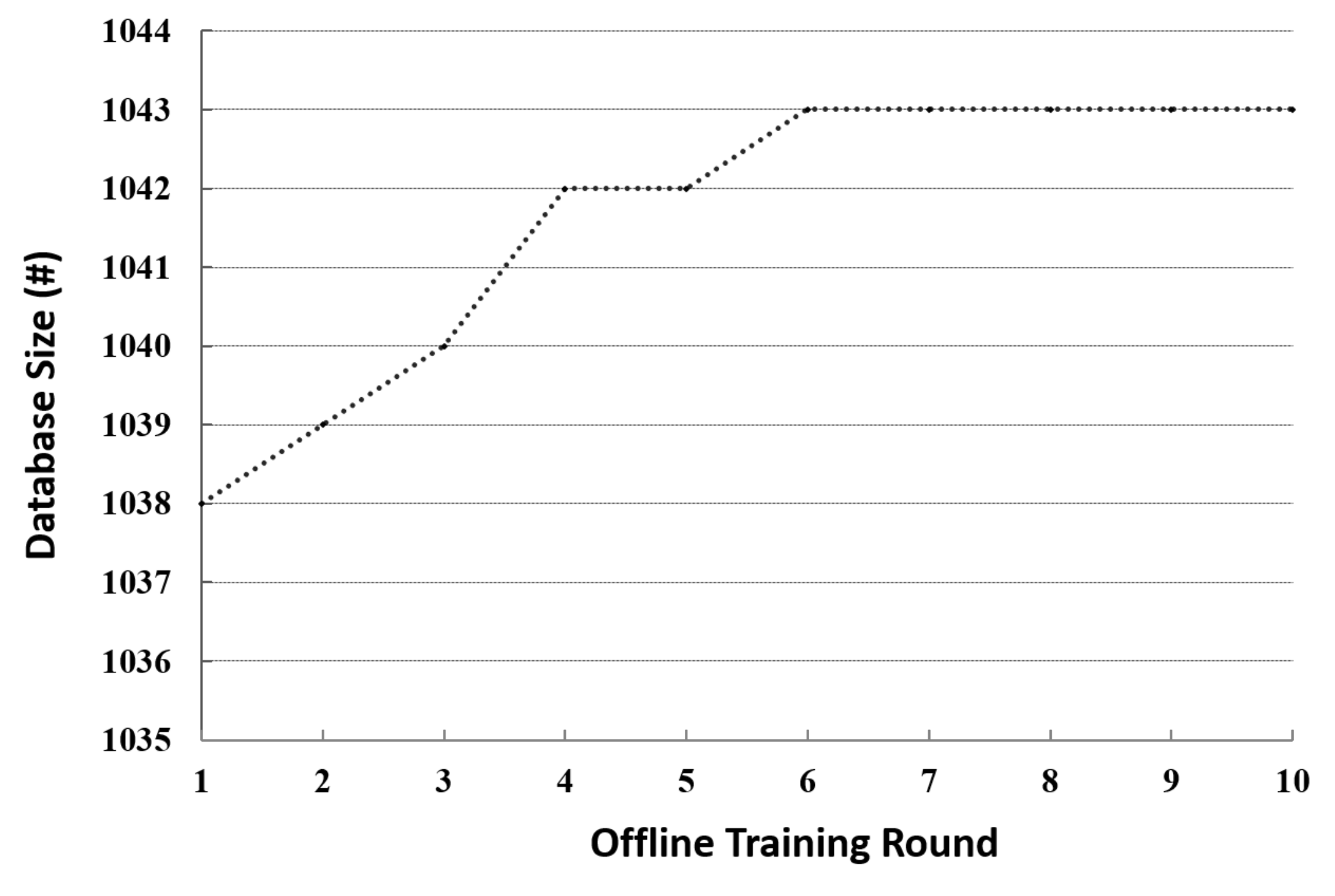}
\caption{In the case of \texttt{LAMP}, its database is built from scratch and keeps its size increasing until the round $6$\emph{th}.}
\label{fig:lamp-db}
\end{minipage}%
\hspace{0.08cm}
\begin{minipage}[t]{0.49\linewidth}
\centering
\includegraphics[width=0.99\linewidth, height=0.3\textheight]{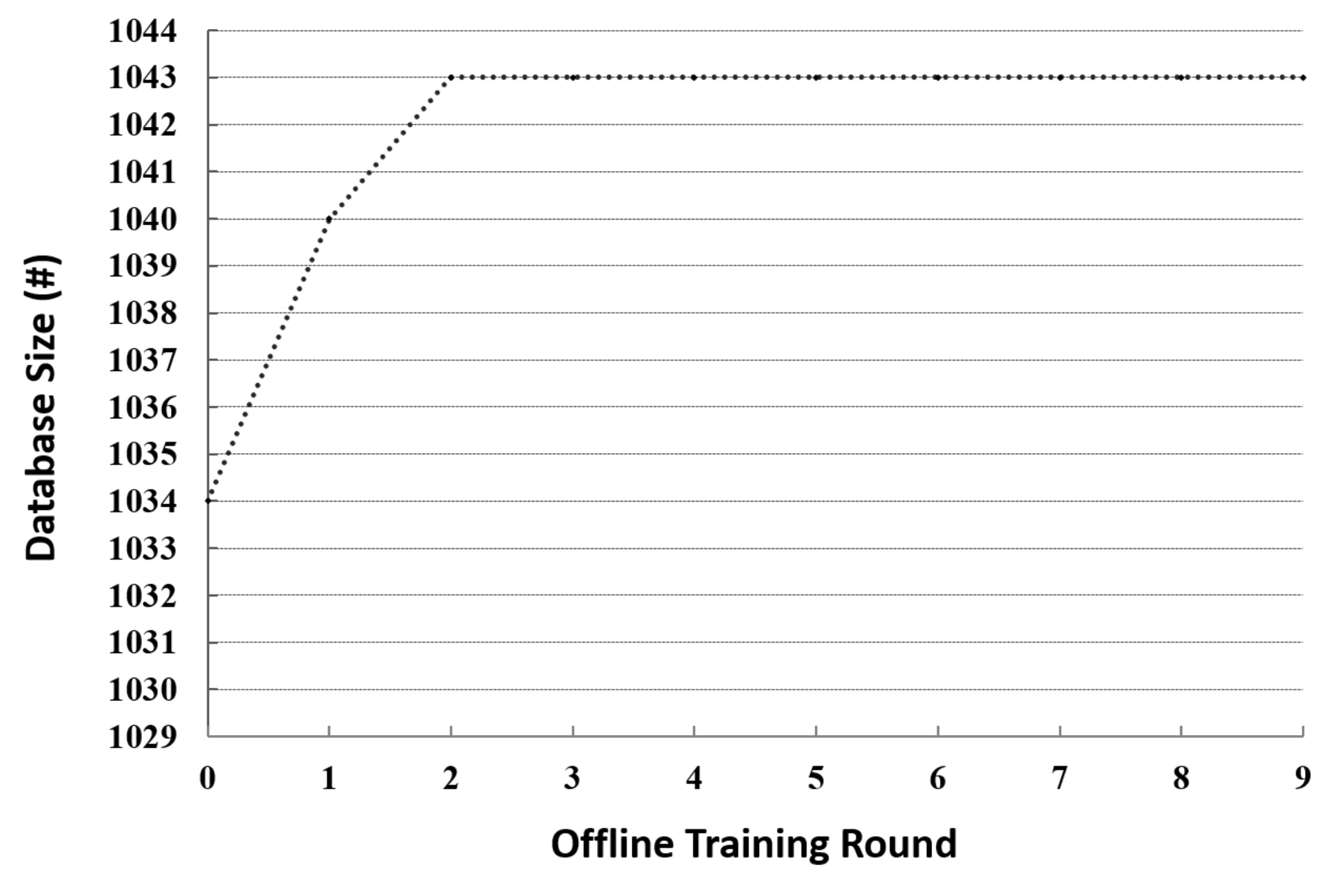}
\caption{\textbf{Incremental offline training.} Compared to that of Figure~\ref{fig:lamp-db}, only $2$ more offline training rounds based on a provided database are needed to reach the same stable state, largely reducing the offline training cost.}
\label{fig:incre-train}
\end{minipage}
\end{figure}

We take \texttt{LAMP} server as an example to illustrate the offline training  efficiency, indicating how fast to construct a stable database for a given workload.
Specifically, we repeat the offline training stage for several rounds to build the \emph{LAMP} database from scratch.
After the first round, we get $1038$ code pages, $99$\% of the final page number.
After that, $9$ successive offline training rounds are completed one by one, each of which updates the database based on previous one, ensuring that the final database records all used pages.
From Figure~\ref{fig:lamp-db}, it can be seen that 
the database as a whole becomes steady after multiple rounds (i.e., $6$ in our experiments).
This observation is also confirmed in other cases.

In fact, it is still time-consuming to build a particular database from scratch. To further accelerate this process, we attempt to do the offline training stage from an existing database.
In our experiments, we integrate every database generated respectively for \texttt{SPECint}, \texttt{httperf}, \texttt{bonnie++} into a larger one, and try to generate the \texttt{LAMP} database using incremental training.
Based on the integrated database, we find that only $2$ rounds are enough to generate the stable database for \texttt{LAMP}, shown in Figure~\ref{fig:incre-train}, significantly improving the offline training efficiency.

\section{Discussion} \label{sec:dis}
In this section, we will discuss limitations of our approach.


\mypara{Training Completeness}
Similar to Ktrim~\cite{Ktrim}, KRAZOR~\cite{kurmus2014quantifiable} and Face-Change~\cite{gu2014face}, \name also uses a training-based approach.
As the approach might miss some corner cases, it may cause \name to mark certain pages that should be used as unused, resulting in an incomplete offline training database.
Theoretically speaking, it is possible for such situations to occur. However, in practice, they have never been observed in our experiments so far.
Interestingly, Kurmus al et.~\cite{attacksurface-ndss} found that a small offline training set is usually enough to cover all used kernel code for a given use case, implying that the corner cases usually do not increase the kernel code coverage. 
If the generated database is incomplete, EPT violations may have been triggered at runtime. For such situations, \name has two possible responses. One is to directly stop the execution of the guest, which is suitable for the security sensitive environment where any violations may be treated as potential attacks. The other one is to generate a log message, which is friendly to the applications that have high availability requirements. The generated log contains the execution context and the corresponding memory content to facilitate a further analysis, e.g., system forensics.

\eat{
\subsection{Portability}
Although its current implementation is based on the x86 platform, the design of \name can be ported to the ARM platform.
Currently, the virtualization technique is available on the ARM platform, and we do not see any technical barrier to the migration.
We even identify several advantages on ARM platform, e.g., the ARM instructions are of equal lengths and thus it naturally avoids the trouble introduced by the variable-length instructions (see Section~\ref{sec:atm}).
Besides, different from existing works in Section~\ref{sec:introduction}, \name essentially requires no kernel source code to achieve an efficient kernel attack surface reduction. 
Thus, it provides a generic approach to enhancing the security of commodity OS kernels (e.g., Windows).
}


%
\mypara{Fine-grained Segmentation}
By default, we have three segmented phases (i.e., \emph{startup}, \emph{runtime}, and \emph{shutdown}). 
Actually, the whole lifecycle could be segmented into more phases, corresponding to different working stages of a user application. Intuitively, a more fine-grained segmentation will achieve a better kernel attack surface reduction. Nonetheless, more phases will introduce more performance overhead, such as the additional phase switches. 
In addition, it will increase the complexity of the \name offline training processor, and consequently increases the trusted computing base (TCB).
At last, the \name module has to deal with the potential security attacks, e.g., malicious phase switches.
To prevent such attacks, a state machine graph of phases should be provided, where the predecessor, successor and the switch condition of each phase should be clearly defined. 
At runtime,  the \name module will load this graph and enforce the integrity: only the phase switches existing in the graph are legal, and any other switches will be rejected. 

\eat{
\subsection{Instruction-level Reduction}
As we can see, if unused instructions and used instructions coexist within the same page, current \name system cannot separate them and has to keep them in the attack surface, due to the coarse-grained (page-level) granularity.
To address the issue, a fine-grained instruction-level approach is expected.
One possible solution is to use binary rewriting~\cite{chanet2005system} to remove the unused instructions from the mixed pages.
Nevertheless, this solution obviously breaks the kernel code integrity, which violates our system goals.
How to achieve the find-grained reduction while follow the system goals is still an open question.

\mypara{Code-reuse/DOP Rootkits}
Besides the code-injection attacks, \name may alleviate other types of kernel rootkits: (1) code reuse attacks, and (2) Data-Oriented Programming attacks~\cite{hu2015automatic}.

For code reuse attacks such as ROP~\cite{prandini2012return} and JOP~\cite{bletsch2011jump}, they do not inject any code, instead they reuse existing code to select useful code blocks (i.e., \emph{gadgets}) to launch attacks. 
As \name is able to reduce/remove more than half of the code pages, the number of gadget candidates will be correspondingly reduced.
Consequently, it raises the bar of launching code reuse attacks due to lack of expected gadgets.

The DOP attack~\cite{hu2015automatic} mainly relies on chained gadgets and gadget dispatchers to launch memory exploits, which strictly conforms to the CFI~\cite{abadi2005control}. Thus, such attacks cannot be detected by CFI.
Fortunately, both the gadgets and the dispatchers are still composed of existing code blocks in the kernel space. 
It is thus possible for \name to alleviate the attack. Demonstrating \name against code-reuse/DOP rootkits is our future work. 
}

\eat{
\subsection{Database Generation}
Except using the dynamic training based approach to generate database, we also investigate the static approaches.

In fact, a static approach is
Alternatively, it is possible for skilled system administrators to combine static and dynamic binary analysis of specific use cases in order to produce such databases, although it is rather difficult. One of the challenges raised by this method is the well-known issue:dynamic bound, which occurs to C++ applications.
In a conclusion, it is highly recommended to apply incremental training when generating a database.
}

\section{Related Work} \label{sec:related}
In this section, we provide an overview of existing approaches to enhance the kernel security that require no changes to the kernel. Specifically, the approaches are either kernel or hypervisor-dependent.

Kernel customizations~\cite{tartler2012automatic,attacksurface-ndss} present automatic approaches of trimming kernel configurations adapted to specific use cases so that the tailored configurations can be applied to re-compile the kernel source code, thus minimizing the kernel attack surface. Similarly, Seccomp~\cite{Seccomp} relies on the kernel source code to sandbox specified user processes by simply restricting them to a minimal set of system calls.
Lock-in-Pop~\cite{li2017lock} modifies and re-compiles \emph{glibc} to restrict an application' access to certain kernel code. In contrast, 
both Ktrim~\cite{Ktrim} and KRAZOR~\cite{kurmus2014quantifiable} utilize \emph{kprobes} to trim off unused kernel functions and prevent them from being executed. All of the approaches above aim at providing a minimized kernel view to a target application.

In the virtualized environment, both Secvisor~\cite{seshadri2007secvisor} and NICKLE~\cite{NICKLE} only protect original kernel TCB and do nothing to reduce it. Taking a step further, 
unikernel~\cite{2013unikernels} provides a minimal kernel API surface to specified applications but developing the applications is highly dependent on the underlying unikernel. 
Face-Change~\cite{gu2014face} profiles the kernel code for every target application and uses the Virtual Machine Introspection (VMI) technique to detect process context switch and thus provide a minimized kernel TCB for each application. 
However, Face-Change has three disadvantages: (1) Its worst-case runtime overhead for \texttt{httperf} testing Apache web server is $40$\%, whereas our worst overhead is $1.94$\% (see Section~\ref{sec:pe}), making it impractical in the cloud environment. (2) Its design naturally does not support KASLR, which is an important kernel security feature and has been merged into the Linux kernel mainline since kernel version 3.14. In contrast, \name is friendly to the security feature. (3) While multiple-vCPU support is critical to system performance in the cloud environment, it only supports a single vCPU within a guest VM, whereas \name allocates four vCPUs to the VM.

\eat{
In this section, we make a comparison of different approaches that reduce the kernel attack surface.
 
As discussed before, in order to reduce the kernel attack surface, existing approaches either build a security-enhanced kernel from scratch (i.e., a new kernel architecture), or re-construct current kernel architecture, or customize an OS kernel.

\textbf{Build From Scratch.}
Designing a kernel as small and modular as possible is always the primary goal for micro-kernel architectures~\cite{accetta1986mach,herder2006construction,herder2006minix,sel4}, among which
Mach~\cite{accetta1986mach} is one of the earliest examples of a micro-kernel, implements a small but extensible system kernel, thus achieving a minimal trusted computing base (TCB). However, compared to a native monolithic UNIX kernel, Mach has serious performance issues. Sel4~\cite{sel4} is proposed as the world's first OS that achieves a high degree of assurance through formal verification. Besides, it works significantly faster than any other micro-kernels on the supported processors. Note that \name works transparently for the guest OS and thus is complementary to this category of approaches. 

\textbf{Re-Construction.}
As device drivers are more buggy or vulnerable than core kernel~\cite{swift2002nooks}, Nooks~\cite{swift2002nooks}, LXFI~\cite{LXFI2011software} and SUD~\cite{boyd2010tolerating} attempt to isolate device drivers to protect the core kernel from being compromised. A common drawback of these techniques is that they do trust the core kernel. On the contrary, Nested Kernel~\cite{nestedkernel-asplos} does not trust the whole monolithic kernel. Instead, it introduces a small isolated kernel, removing the monolithic kernel from TCB. Alternatively, there are efforts made to enforce access control within the monolithic kernel~\cite{cook2013linux,liakh2010analyzing,selinux2001implementing} or restrict the use of system calls~\cite{Seccomp}.

\textbf{Customization.}
This category will not make any changes to the kernel and 
}

\section{Conclusion} \label{sec:con}
Commodity OS kernels provide a large number of features to satisfy various demands from different users, exposing a huge surface to remote and local attackers.
In this paper, we have presented a \emph{reliable} and \emph{practical} approach, named \name, which has \emph{transparently} reduced attack surfaces of commodity OS kernels at runtime without relying on their kernel source code.
\name deploys two surface reduction approaches. One is spatial, i.e., the \emph{permission deprivation} marks never-used code pages as non-executable while the other is temporal, i.e., the \emph{lifetime segmentation} selectively activates appropriate used code pages.
We implemented \name on the Xen hypervisor and evaluated it using the Ubuntu OS with an unmodified Linux kernel.
The experimental results showed that \name has efficiently reduced $64\%$ of kernel attack surface, $40\%$ of CVEs in all given use cases. In addition, \name defeated all $6$ real-world rootkits and incurred low performance overhead (i.e., less than $1\%$ on average) to the whole system.

In the near future, our primary goals are to apply \name to the kernel attack surface reduction of a Windows OS since \name should be generic to protect all kinds of commodity OS kernels.

%





	\bibliography{kasr}
\end{document}